\documentclass[aps,prd,preprint,nofootinbib,groupedaddress]{revtex4-1}
\usepackage{graphicx}% Include figure files
\usepackage{dcolumn}% Align table columns on decimal point
\usepackage{bm}% bold math
\usepackage{amsmath,amssymb}

\usepackage{color}

\begin{document}

\preprint{CYCU-HEP-16-07}

\title{Holographic Model of Dual Superconductor for Quark Confinement}% Force line breaks with \\

\author{Tsung-Sheng Huang}\thanks{
E-mail: r04222035@ntu.edu.tw}
\affiliation{%
 National Taiwan University, Taipei, 10617 Taiwan
}%
\author{Wen-Yu Wen}\thanks{
E-mail: steve.wen@gmail.com}
\affiliation{%
Department of Physics and Center for High Energy Physics,
 Chung Yuan Christian University, Chung Li District, Taoyuan
}%
\affiliation{Leung Center for Cosmology and Particle Astrophysics\\
National Taiwan University, Taipei 106, Taiwan}

\begin{abstract}

We show that a hairy black hole solution can provide a holographically {\sl dual} description of quark confinement.   There exists a one-parameter sensible metric which receives the backreaction of matter contents in the holographic action, where the scalar and gauge field are responsible for the condensation of chromomagnetic monopoles.  This model features a preconfining phase triggered by second-order monopole condensation and a first-order confinement/deconfinement phase transition.  To confirm the confinement, the quark-antiquark potential is calculated by probing a QCD string in both phases.  At last, contribution from Kaluza-Klein monopoles in the confining phase is discussed.

\end{abstract}

\pacs{Valid PACS appear here}% PACS, the Physics and Astronomy
\maketitle

%\tableofcontents

%\section{\label{sec:level1}Introducion}

The holographic correspondence between a gravitational theory in the bulk and a quantum field theory on the boundary, first emerged under the anti-de Sitter/conformal field theory (AdS/CFT) correspondence \cite{Maldacena:1997re,Gubser:1998bc,Witten:1998qj}, has been proved useful to study condensed matter phenomena.  In particular, the authors in \cite{Gubser:2005ih,Gubser:2008px} proposed a gravity model in which Abelian symmetry of Higgs is spontaneously broken by the existence of black hole, also known as hairy black hole.  This mechanism was incorporated in the model of superconductivity and critical temperature was observed \cite{Hartnoll:2008vx,Gubser:2008zu}.  This model was later studied in the presence of magnetic field \cite{Nakano:2008xc,Albash:2008eh} and with full backreaction \cite{Hartnoll:2008kx}.

On the other hand, the AdS/CFT analogy has also explained many qualitative and quantitative features of confining gauge theories like QCD, either by engineering branes and strings in the ten dimensional string theory \cite{Kruczenski:2003uq,Sakai:2004cn}, or by bottom-up construction of a Randall-Sundrum like background \cite{Erlich:2005qh,DaRold:2005mxj}.  In particular, holographic models of deconfinement were first discussed in the Sakai-Sugimoto model at finite temperature\cite{Aharony:2006da} and the hard/soft wall model\cite{Herzog:2006ra}.   In all cases, the confinement/deconfinement transition is identified as the first-order Hawking-Page phase transition, where the thermal AdS is energetically favored at low temperature, while a black hole solution at high temperature \cite{Hawking:1982dh}.  In this paper we propose the first holographic confinement model without using the Hawking-Page phase transition.

We recall an old but appealing proposal of dual superconductor \cite{Nambu, tHooft, Mandelstam}, stating that the confining phase can be regarded as a dual type II superconducting state due to the condensation of chromomagentic charges (monopoles), such that the chromoelectric lines between a quark-antiquark pair are squeezed by dual Meissner effect into flux tubes, as known as the QCD strings.   One may wonder if the hairy black hole scenario in the holographic model of dual superconductor\cite{Hartnoll:2008vx} works just fine by trivially replacing Cooper pairs by monopole condensate, and the backreacted metric in the \cite{Hartnoll:2008kx} might be the confinement geometry in desire.  However, a second thought and straightfoward computation will show this is not the case.  The backreaction of condensate may have resulted to a gapped excitation as shown in the single electron spectral function \cite{Chen:2009pt,Faulkner:2009am,Gubser:2009dt}, but the attractive force mediated by phonons is coulomb-like, rather than a linear potential for flux tube in the dual picture.  Then the challenging task is to find out the appropiate ansatze which can produce the linear potential while condensation occurs.   In this letter, we consider a $(2+1)$-dimensional dual supercondcutor model which has a holographic description of Eisntein-Maxwell-dilaton gravity theory in the AdS$_4$.  The action is composed of a gravity sector $S_g$, a matter sector $S_M$ for massive scalars and a dual $U(1)$ gauge field sector $S_A$ as follows:
\begin{eqnarray}
&&S = S_g + S_M + S_A,\nonumber\\
&&S_g = \int{dtdzd\vec{x}} ~\{R + \frac{6}{L^2} \}, \nonumber\\
&&S_M = \int{dtdzd\vec{x}} ~\{ -|\partial \psi -iqA |^2 + m^2|\psi|^2 \}, \nonumber\\
&&S_A = \int{dtdzd\vec{x}} ~\{-\frac{1}{4}F^2\}.
\end{eqnarray}

We will study a class of metric of following general form:
\begin{align}\label{ansatz}
ds^2=-g_{0}(z) dt^2+g_{1}(z) dz^2 +g_{2}(z) d\vec{x}^2
\end{align}
with 

\begin{align}
&g_{0}(z)=\frac{exp(\alpha (p  z/z_{h})^2)}{z^2}[1-(\frac{z}{z_{h}})^3]\nonumber\\
&g_{1}(z)=\frac{exp(\beta (p z/z_{h})^2)}{z^2}\frac{1}{1-(\frac{z}{z_{h}})^3}\nonumber\\
&g_{2}(z)=\frac{exp(\gamma (p z/z_{h})^2)}{z^2}
\end{align}

\begin{figure}
\centering
\includegraphics[width=80mm]{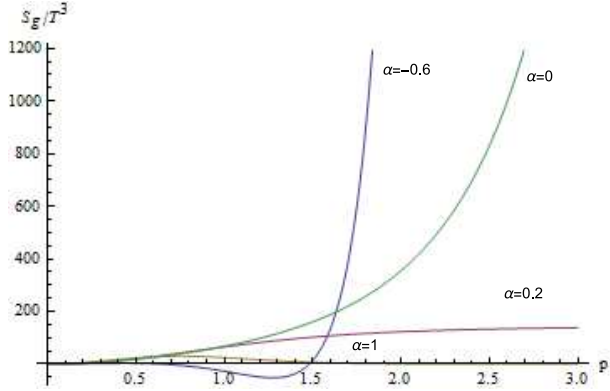}
\caption{The onshell action of pure gravity sector versus $p$ for different $\alpha$. The blue, green, purple, yellow curves correspond to $\alpha=$-0.6, 0, 0.2 and 1, respectively. It is obvious that the blue ($\alpha=-0.6$) and yellow ($\alpha=1$) curves are undesirable since that they have additional local minimum at some $p$ other than $0$.} 
\label{alpha_value} %label figure
\end{figure}

The metric corresponds to a hairy black hole with a flat horizon at $z=z_h$ and approaches asymptotically to AdS as $z\to 0$.  One can show by explicit calculation that the oversimplified ansatz $\beta=\gamma=0$, which was adopted in the backreacted holographic superconductor such as one in \cite{Hartnoll:2008kx}, is unable to produce the linear potential in the confining phase.  As shown in the appendix, the warping factors can be further pinned down to $\beta=\frac{\alpha-2}{3}$ and  $\gamma=1-\alpha$ after imposing the condition that the onshell action of gravity sector is independent of the choice of $p$ after regularization.  The total free energy is given by summing up onshell action of all sectors and the value of $p$ is determined by minimizing the total free energy at a given temperature.  The variable $\alpha$, on the other hand, is a free parameter, which can be fine tuned to match the tension of flux tube.   In practice, the desirable onshell action of gravity sector $S_g$ should have local minimum only at $p=0$, which happens as $\alpha$ is within some range as shown in the Figure \ref{alpha_value}.

In the deconfining phase, the solution (\ref{ansatz}) reduces to the AdS Schwarzschild black hole, while confining parameter $p$ is zero and the oneshell actions of scalar and gauge sectors are trivially zero. 

As temperature decreases, one expects a transition to the confining phase as the the black hole develops hairs after receiving backreaction from condensate.   The scalar field, which represents the order parameter of chromomagnetic charge condensate, poccesses the asymptotical form: 

\begin{figure}[h]
\centering
\includegraphics[width=80mm]{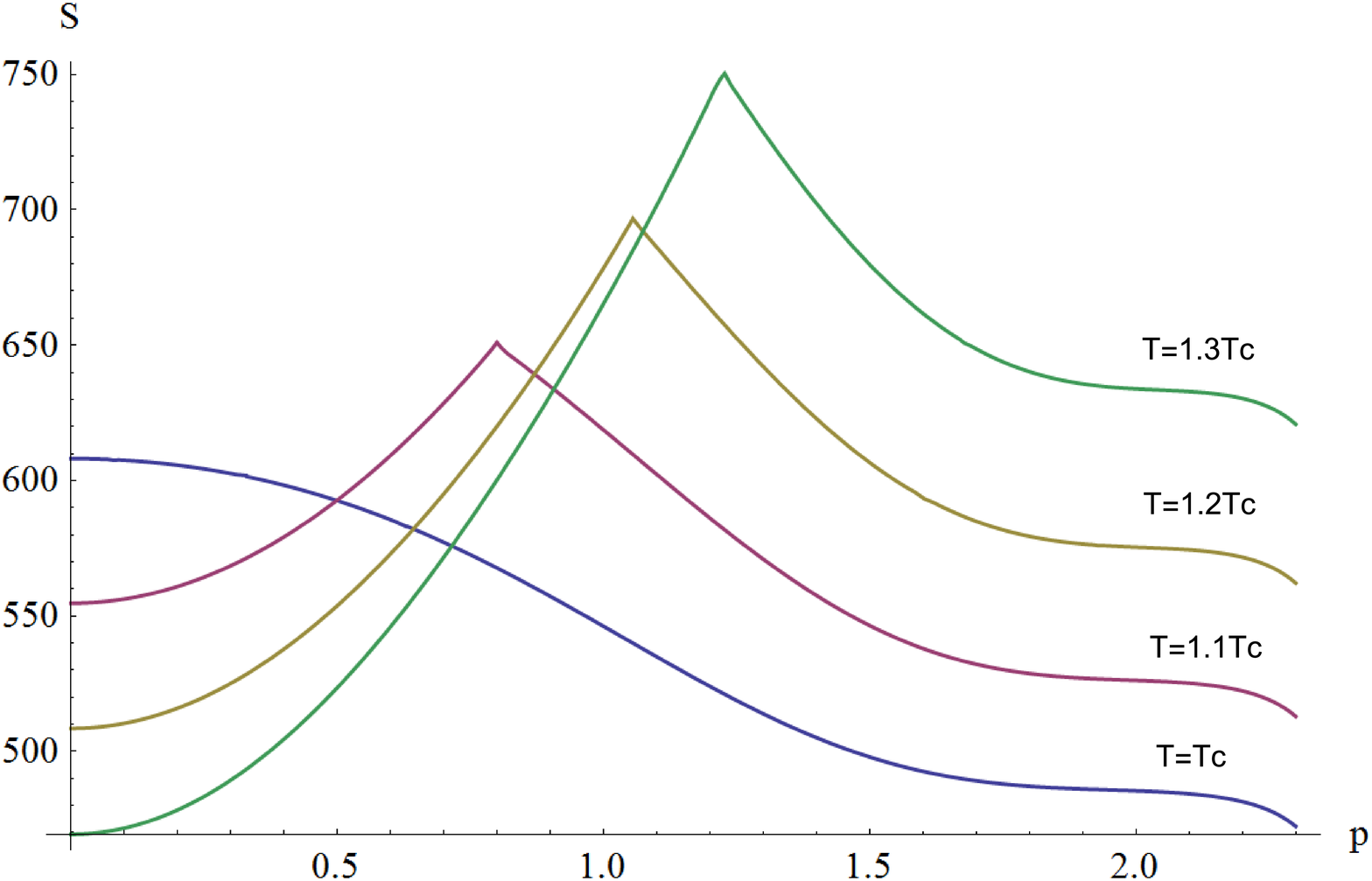}
\caption{The relation between the total onshell action $S$ and confining parameter $p$ at different temperatures. Those curves associated to temperature above $T_c$ usually exhibit a cusp at $p=p_{m}$ and a cutoff at $p_{M}\simeq 2.344$, in between the monopole condensation occurs. The blue, purple, yellow and green curves correspond to the scaled temperatures $T/T_c$ at 1.0, 1.1, 1.2, 1.3, respectively.  One observes competition between local minima $p=p_M$ (confining phase) and $p=0$ (deconfining phase) at the temperature $T_c^*$, which is between $1.1T_c$ and $1.2T_c$, signaling a preconfining phase.  There is only a confining phase when temperature is below $T_c$.  Here we set $m^2=-2$, $\rho=71.5$ and $\alpha=0.2$ for simulation. } %add the caption
\label{onshell_action} %label figure
\end{figure}

\begin{align}
\psi={\cal{S}} z^{\Delta_{-}}+{\cal{C}}z^{\Delta_{+}}+\cdots,
\end{align}
with $\Delta_{\pm}=\frac{3\pm\sqrt{9+4m^2}}{2}$, where ${\cal{S}} $ and ${\cal{C}} $ are interpreted as source and condensate, respectively.     We remark that one can also switch the role of source and condensate for a scalar field in asymptotical AdS$_4$.  That would correspond to the condensate of operators with scaling dimension $\Delta_{-}$.  The chemical potential $\mu$ and charge density $\rho$ can be read out from the asymptotic form of gauge field:

\begin{align}
\phi=\mu-\rho z +{\cal{O}}(z^2)
\end{align}
In our simulation, the same boundary conditions adopted in the \cite{Hartnoll:2008vx} are imposed, that is, ${\cal S}=0$ and $\rho$ is fixed at the asymptotic boundary and $\phi=0$ at the horizon.  The condensate ${\cal C}$ and chemical potential $\mu$ are obtained numerically by the shooting methods.
The phase transition is justified by evaluation of the total onshell action against confining parameter $p$ at different temperatures.  In the figure \ref{onshell_action}, we see that the confining phase (local minimum at $p=p_M$) is energetically favored at and below the critical temperture $T_c$.  The preconfining phase appears when two local minima $p=0$ and $p=p_M$ competes with each other during $T_c<T<T_c^*$.  The confinement/deconfinement phase transition occurs around $T_c^*$ when the local minimum $p=0$ starts to dominate.  Each onshell curve with temperature above $T_c$ exhibits a cusp at $p=p_m$ and a cutoff.   The cusp is generated by nontrivial condensate ${\cal C}$, which occurs when  $p\geq p_m$.   There is no ground state solution for the scalar $\phi$ above the cutoff.  Since confinement appears only when monopoles condensate, one can conclude that the former phase transition is directly triggered by the latter in the dual superconductor model.   Our model shows that $1.1 T_c<T_c^* < 1.2 T_c$ and predicts the existence of preconfining phase with a fraction window $0.09\leq |T_c-T_c^*|/T_c^* \leq 0.17 $.  We remark that the preconfing phase did not appear in the previous model\cite{Herzog:2006ra} and our result agrees with observation in some nonperturbative approach to thermal $SU(2)$ Yang-Mills theory \cite{Hofmann:2005dt}.  Although the dual superconductor model is the an effective Ginzburg-Landau theory which describes a second-order phase transition for the monopole condensate, we argue that the confinement/deconfinement phase transition in our hairy black hole model is of first order.  This is because temperature in our model is not only determined by the horizon but also the hair, that is

\begin{align}
T=\frac{\sqrt{f_{0}(z_{h})f_{1}(z_{h})}}{4\pi}=\frac{3}{4\pi z_{h}}e^{\frac{2(1+\alpha)p^2}{3}},
\end{align}

The sudden shift of parameter $p$ at the transition causes a discontinuous jump of horizon position $z_h$ and therefore a first-order phase transition.

\begin{figure}[h]
\centering
\includegraphics[width=80mm]{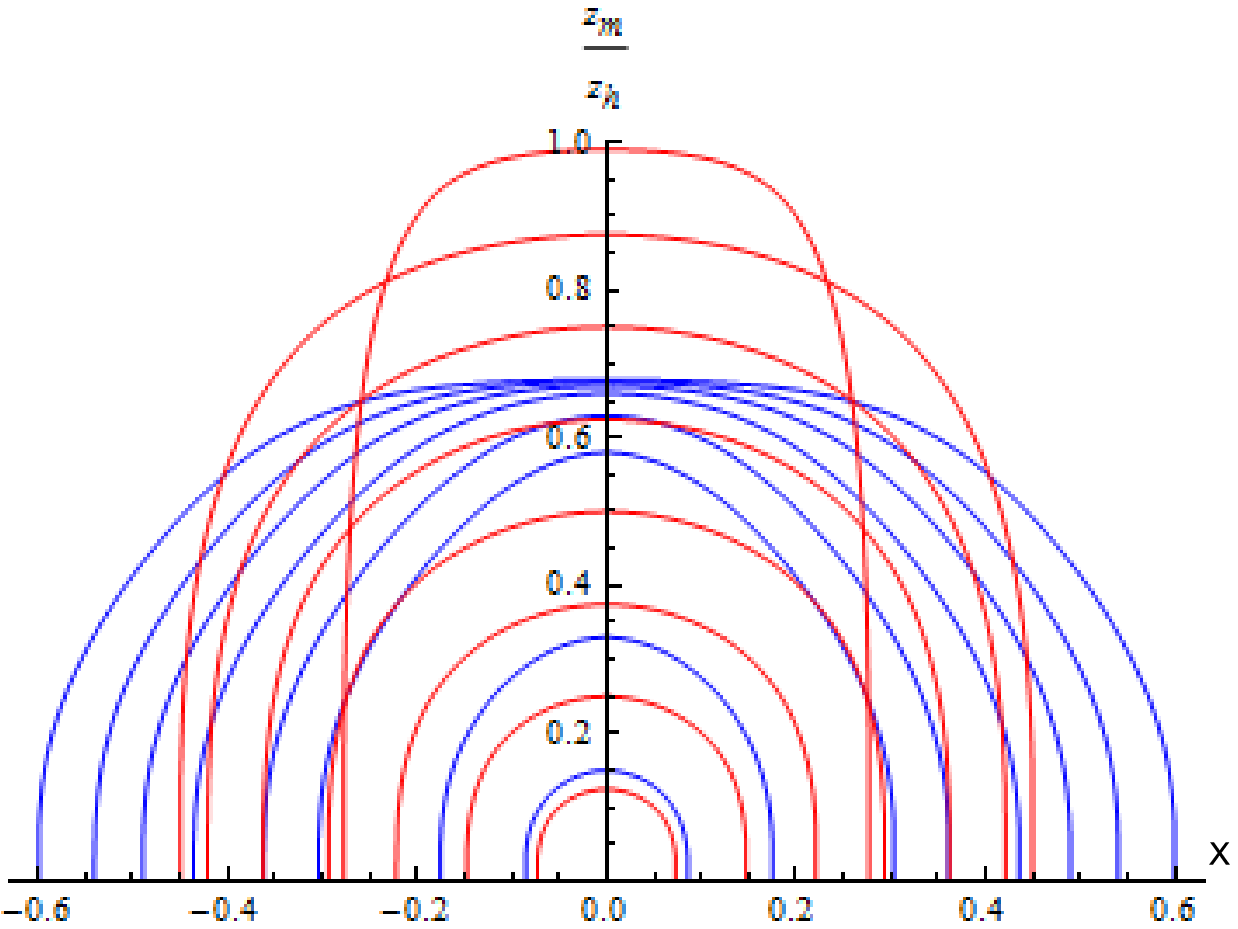}
\includegraphics[width=80mm]{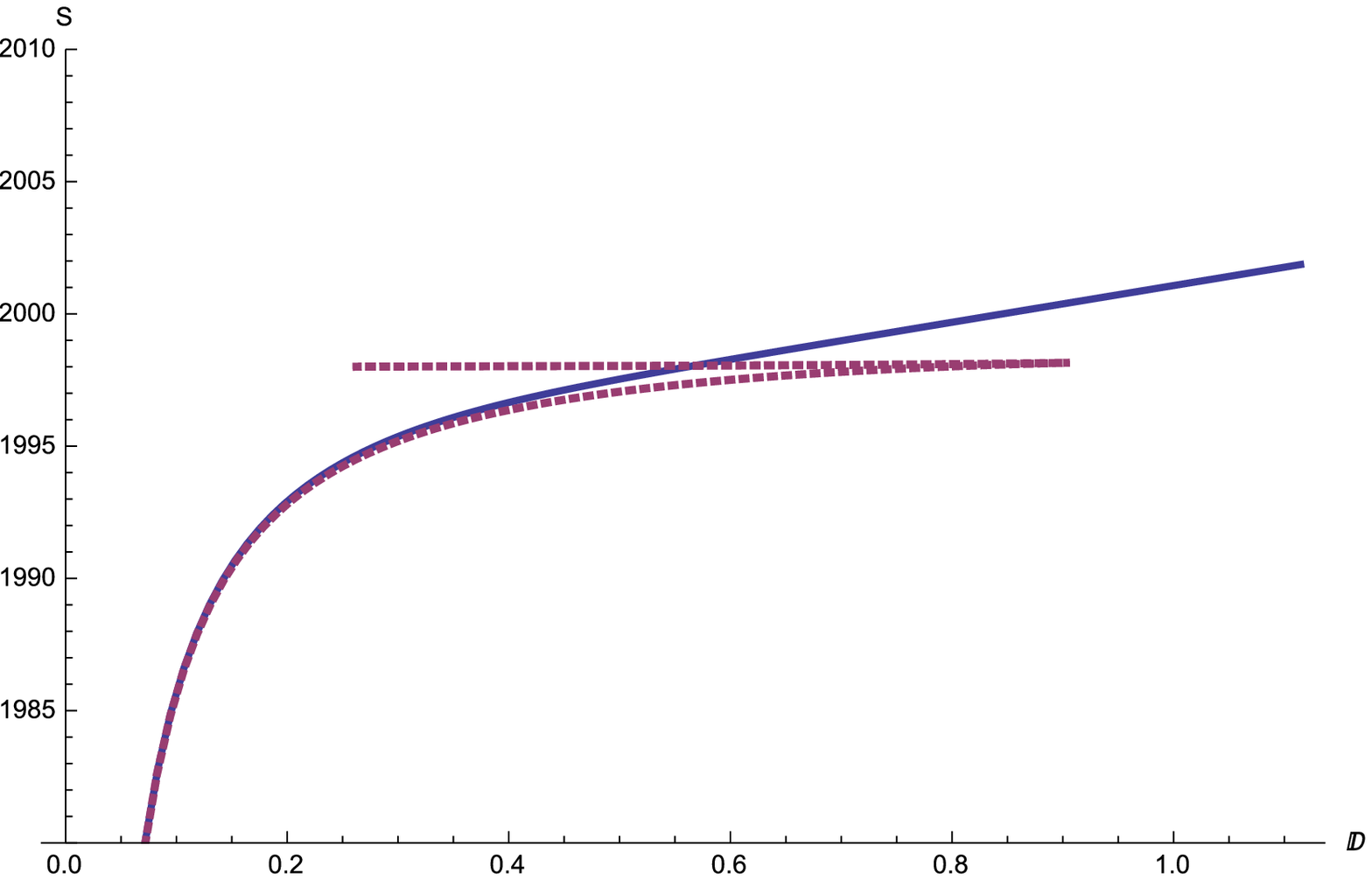}
\caption{(Left) The U-shaped string can reach the horizon in the deconfinement phase (dotted red) but gets expelled from infared core in the confinement phase (solid blue).  (Right) The string onshell action $S_{str}$ is plotted against the seperation between string end points on the boundary.  In the static configuration, the action can be regarded as the potential between a heavy quark-antiquark pair.  In the deconfinement phase, the Coulomb potential (red, lower branch) is terminated as the string is {\sl sucked} into the black hole (red, upper branch).   In the confinement phase, the linear potential (blue) dominates as seperation distance increases. } %add the caption
\label{string} %label figure
\end{figure}

In the imhomogenous background where monopole condensate occurs locally, one may expect both confinement and deconfinement coexist in different regions and the color force lines are squeezed thanks to the dual meissner effect.

One can further confirm both the confinement and deconfinement background by probing a string of profile $z(x)$, given by the Nambu-Goto action

\begin{align}\label{string_action}
S_{str}=\int{\sqrt{|det g_{\mu\nu}|}dx}=\int{\sqrt{g_{0}(g_{2}+(\dot{z})^2g_{1})} dx}
\end{align}

\begin{figure}[h]
\centering
\includegraphics[width=100mm]{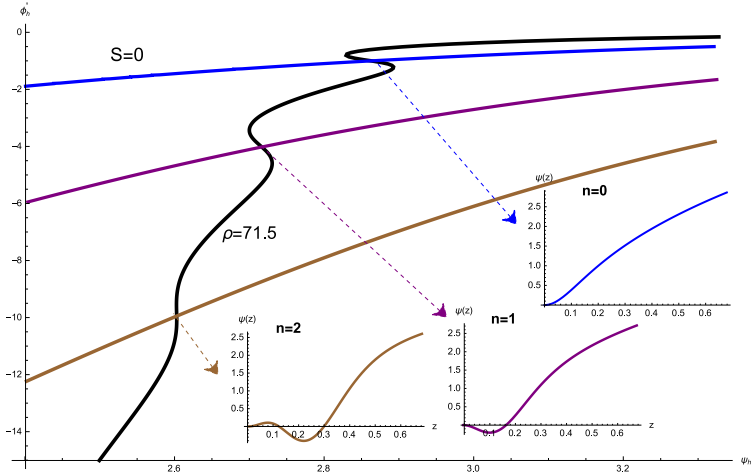}
\caption{The plot demonstrates in general there exist more than one solution in the condensate phase.  These solutions are intersecting points by contour curves of constant $S$ and $\rho$ (black curve) in the $\psi$-$\phi$ plane.  By calculating the wavefunction $\psi_n(z)$ at each intersecting point, we find they correspond to the ground state ($n=0$, blue curve), first excited state ($n=1$, pruple curve), and second excited state ($n=2$, brown curve), respectively.  From viewpoint of boundary theory, those excited modes are the Kaluza-Klein (KK) monopoles with heavier mass.  We use contour curves $S=0$ and $\rho=71.5$, parameters $\alpha=0.2$, and $T=0.3517$ for this simulation.} %add the caption
\label{excitemode} %label figure
\end{figure}

The string profile in the gravity bulk is plotted against the seperating distance in the left figure \ref{string}.  With increasing seperation, the U-shaped string eventually touches the horizon and breaks in the presence of hairless black hole, indicating the deconfinement phase.  In the confining phase, the string can not enter the infared core of bulk in the presence of hairy black hole and stay near $z\simeq  0.68 z_h$.  The horizontal segment of string is responsible to the linear potential.  We remark that string in the confining phase cannot break without introducing additional dynamic degrees of freedom.  The onshell string action can be regarded as the free energy or potential of static heavy quark-antiquark pair, which is plotted against the seperation distance $D$ in the right figure \ref{string}.   In the deconfining phase, though it exhibits a Coulomb potential for small seperation, no force exists for seperation $D>0.9$ in the unit $L=1$  because the string is eventually pulled into the black hole.  In the confining phase, on the other hand, a linear potential has already manifested for seperation $D>0.4$.  

\begin{figure}[h]
\centering
\includegraphics[width=80mm]{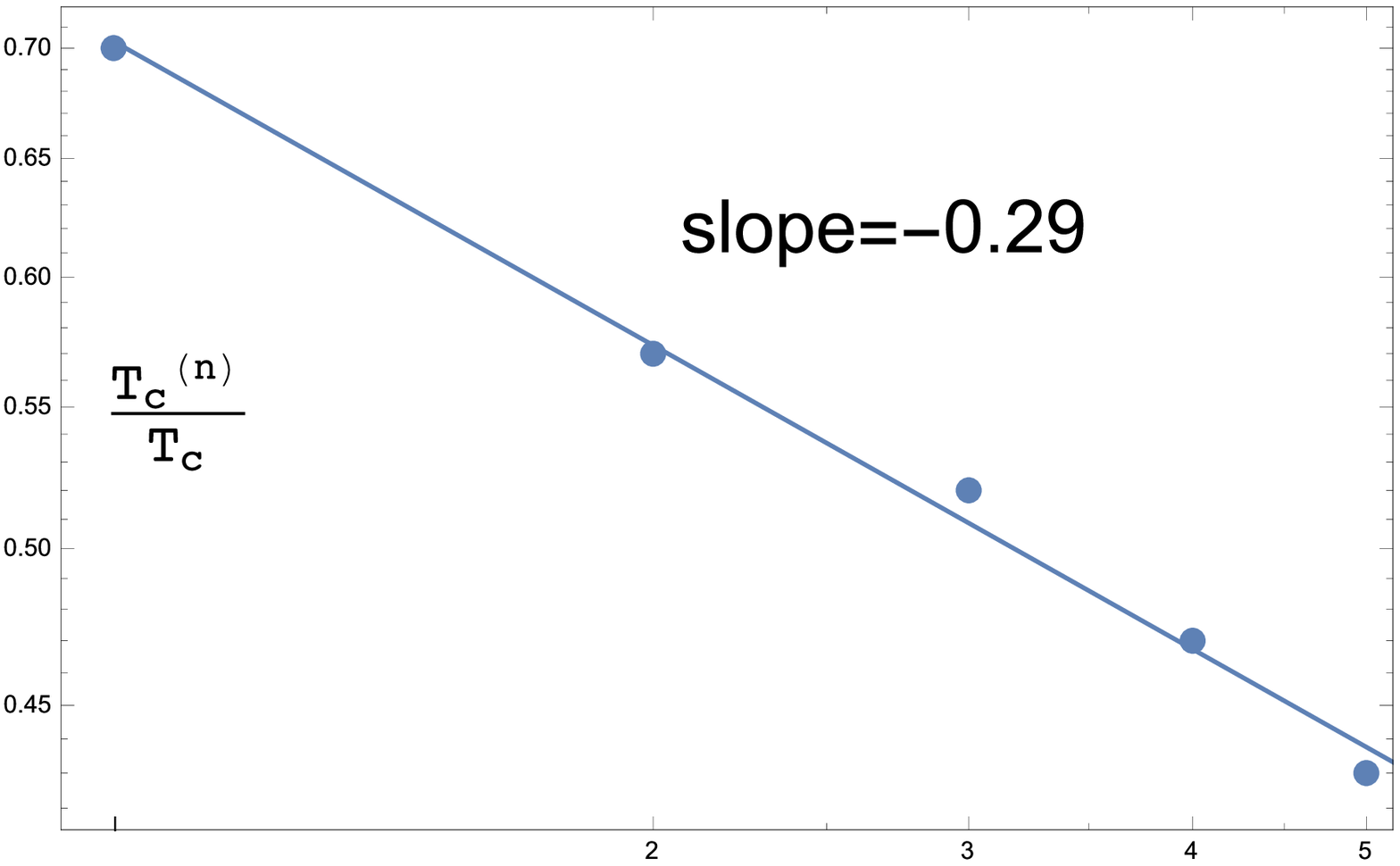}
\includegraphics[width=80mm]{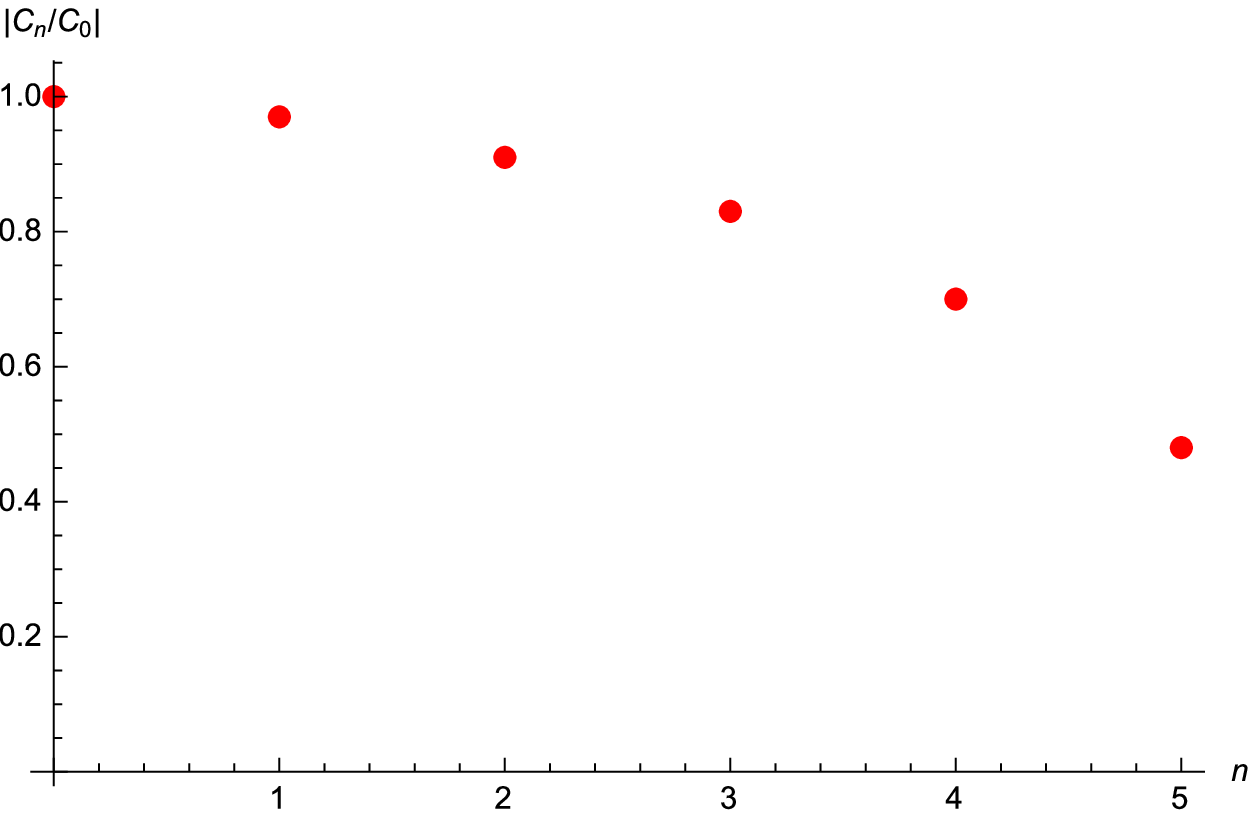}
\caption{(Left) The log plot on the right shows the critical temperature (normalized by that of the ground state) for KK monopole to condensate.  The normalized critical temperature for $n$-excited state can be approaximated by the relation $\frac{T_c^{(n)}}{T_c} \simeq 0.7 n^{-0.29}$. (Right) The plot on the right shows the condensate ratio ${\cal C}_n/{\cal C}_0$ for each KK monopole} %add the caption
\label{monopole} %label figure
\end{figure}

At last, we would like to address a possible role played by excited modes of scalar hair in the confining phase.  As demonstrated in the figure \ref{excitemode}, multiple solutions exist as temperature well below the critical temperature of condensation.  Those solutions correspond to those wavefunctions $\psi_n(z)$ with $n$ nodes, or excited states.  From the point of view of AdS bulk, the tower of excited modes might be related to the Efimov states due to loss of conformality \cite{Faulkner:2009wj,Kaplan:2009kr,Wen:2010et}.  From viewpoint of dual theory on the boundary, those excited states are the Kaluza-Klein (KK) monopoles with effective mass which grows with mode number $n$.  In the figure \ref{monopole}, we plot the critical temperature $T_c^{(n)}$ and condensate ${\cal C}_n$ for the first five KK modes.  We observe that the a $n$-order KK monopole can only be excited at temperature lower than $T_c^{(n)}$.  The measured monopoles mass $m_M$ would be proportional to the expectation value of condensate ${\cal C}_n$ from all avaliable KK modes up to $n^*$ at some temperature, that is $m_M \propto \sum_{n=0}^{n^* } \rho_n {\cal C}_n$.  As a result, an increase in monopole mass $m_M$ is expected as temperature moves away from the critical point in the confining phase thanks to the growing $n^*$.    We remark that the KK monopole was condisered as part of a magnetic bion condensation in some QCD-like theory\cite{Unsal:2007jx}.

In conclusion, we are among the first to construct a holographic model of quark confinement by chromomagnetic monopole condensation.  This toy model features a second-order monopole condensation as well as first-order confinement/deconfinement phase transition, and it predicts a preconfining phase, which is mostly unseen in the previous holographic construction but observed in other nonperturbative approach.  Finally we hypothesize that excited modes of condensate might contribute to the increasing monopole mass in the confining phase.  Our noval result has provided supportive evidence for the application of holographic method and serves as a toy model toward the quark confinement in realistic QCD.

\section*{acknowledgement}

The authors are grateful to the useful discussion with Hsien-Chung Kao and Shoichi Kawamoto in the early stage of this work. WYW would like to thank the warm hospitality received in the USTC, China when part of this work was presented in the East Asia Joint Workshop on Fields and Strings and insightful discussion with Sang-Jin Sin. This work is supported in parts by the Chung Yuan Christian University, the Taiwan's Ministry of Science and Technology (grant No. 102-2112-M-033-003-MY4) and the National Center for Theoretical Science.

\appendix

\section{\label{sec:level1}Warping factors in the background metric}

In this appendix, we will derive the relation among parameters $\alpha,\beta, \gamma$ in the warping factors:

\begin{align}\label{metric_ansatz}
&g_{0}(z)=\frac{exp(\alpha (p  z/z_{h})^2)}{z^2}[1-(\frac{z}{z_{h}})^3]\nonumber\\
&g_{1}(z)=\frac{exp(\beta (p z/z_{h})^2)}{z^2}\frac{1}{1-(\frac{z}{z_{h}})^3}\nonumber\\ 
&g_{2}(z)=\frac{exp(\gamma (p z/z_{h})^2)}{z^2}
\end{align}

At first, we observe the static U-shaped string profile (\ref{string_action}) can be integrated over the differential

\begin{align}
\dot{z}=\sqrt{\frac{g_{2}(z)}{ g_{1}(z)}\left( \frac{h(z)}{h(z_{m})}-1\right) },
\end{align}

providing the top of string locates at $z=z_m$.  Since the function $h(z)\equiv g_{0}(z)g_{2}(z)$ poccesses the following form

\begin{align}
h(z)=\frac{exp((\alpha+\gamma) (p z/z_{h})^2)}{z^4}[1-(\frac{z}{z_{h}})^3],
\end{align}

one can freely absorb the combination $\alpha+\gamma$ into redefinition of $p$, that is to set $\alpha+\gamma=1$.  Furthermore one can eliminate $\beta$ by a choice of regularization scheme, such that all the divergent terms to be {\sl $p$-independent} in the onshell action of gravity sector.  In the other words, one would use the same counterterm in both confinement and deconfinement phases.  To be specific, the onshell action reads

\begin{align}\label{on_shell_action_g}
S_{g}=\int_{\epsilon}^{z_{h}}{L_{g}dz}=\int_{\epsilon}^{z_{h}}{\sqrt{g_{0}(z)g_{1}(z)(g_{2}(z))^2}(R(z)+6)dz}=\frac{2}{\epsilon^3}-\frac{(2-\alpha+3\beta) p^2}{z_{h}^2 \epsilon}+{\cal{O}}(\epsilon^{0}),
\end{align}
in the unit of $L=1$, where the Lagrangian density can be calculated explcitly:
\begin{widetext}
\begin{align}
L_{g}=&\frac{e^{-\frac{p^2 z^2 (\alpha +\beta -2)}{2 z_h^2}}}{z^2 z_h^2}\left(p^2 \left((\alpha +3 \beta +4) \left(\left(\frac{z}{z_h}\right)^3-2\right)-2 (\alpha -8)\right)\right)+\frac{6 e^{-\frac{p^2 z^2 (\alpha +\beta -2)}{2 z_h^2}}}{z^4}\left(e^{\frac{\beta  p^2 z^2}{z_h^2}}-2\right) \nonumber\\
&+\frac{e^{-\frac{p^2 z^2 (\alpha +\beta -2)}{2 z_h^2}}}{z_h^4}\left(2 p^4 \left(2 \alpha ^2+\alpha  (\beta -4)-2 \beta +3\right) \left(\left(\frac{z}{z_h}\right){}^3-1\right)\right).
\end{align}
\end{widetext}

To make the second term in (\ref{on_shell_action_g}) vanish for arbitrary $p$, one can enforce $\alpha+3\beta+2\gamma=0$, which gives another constraint to eliminate $\beta$.


\begin{thebibliography}{99}

%\cite{Maldacena:1997re}
\bibitem{Maldacena:1997re} 
  J.~M.~Maldacena,
  %``The Large N limit of superconformal field theories and supergravity,''
  Int.\ J.\ Theor.\ Phys.\  {\bf 38}, 1113 (1999)
  [Adv.\ Theor.\ Math.\ Phys.\  {\bf 2}, 231 (1998)]
  doi:10.1023/A:1026654312961
  [hep-th/9711200].
  %%CITATION = doi:10.1023/A:1026654312961;%%
  %11793 citations counted in INSPIRE as of 18 May 2016

%\cite{Gubser:1998bc}
\bibitem{Gubser:1998bc} 
  S.~S.~Gubser, I.~R.~Klebanov and A.~M.~Polyakov,
  %``Gauge theory correlators from noncritical string theory,''
  Phys.\ Lett.\ B {\bf 428}, 105 (1998)
  doi:10.1016/S0370-2693(98)00377-3
  [hep-th/9802109].
  %%CITATION = doi:10.1016/S0370-2693(98)00377-3;%%
  %6734 citations counted in INSPIRE as of 18 May 2016

%\cite{Witten:1998qj}
\bibitem{Witten:1998qj} 
  E.~Witten,
  %``Anti-de Sitter space and holography,''
  Adv.\ Theor.\ Math.\ Phys.\  {\bf 2}, 253 (1998)
  [hep-th/9802150].
  %%CITATION = HEP-TH/9802150;%%
  %7771 citations counted in INSPIRE as of 18 May 2016

%\cite{Gubser:2005ih}
\bibitem{Gubser:2005ih} 
  S.~S.~Gubser,
  %``Phase transitions near black hole horizons,''
  Class.\ Quant.\ Grav.\  {\bf 22}, 5121 (2005)
  doi:10.1088/0264-9381/22/23/013
  [hep-th/0505189].
  %%CITATION = doi:10.1088/0264-9381/22/23/013;%%
  %116 citations counted in INSPIRE as of 18 May 2016

%\cite{Gubser:2008px}
\bibitem{Gubser:2008px} 
  S.~S.~Gubser,
  %``Breaking an Abelian gauge symmetry near a black hole horizon,''
  Phys.\ Rev.\ D {\bf 78}, 065034 (2008)
  doi:10.1103/PhysRevD.78.065034
  [arXiv:0801.2977 [hep-th]].
  %%CITATION = doi:10.1103/PhysRevD.78.065034;%%
  %617 citations counted in INSPIRE as of 18 May 2016

%\cite{Hartnoll:2008vx}
\bibitem{Hartnoll:2008vx} 
  S.~A.~Hartnoll, C.~P.~Herzog and G.~T.~Horowitz,
  %``Building a Holographic Superconductor,''
  Phys.\ Rev.\ Lett.\  {\bf 101}, 031601 (2008)
  doi:10.1103/PhysRevLett.101.031601
  [arXiv:0803.3295 [hep-th]].
  %%CITATION = doi:10.1103/PhysRevLett.101.031601;%%
  %903 citations counted in INSPIRE as of 18 May 2016

%\cite{Gubser:2008zu}
\bibitem{Gubser:2008zu} 
  S.~S.~Gubser,
  %``Colorful horizons with charge in anti-de Sitter space,''
  Phys.\ Rev.\ Lett.\  {\bf 101}, 191601 (2008)
  doi:10.1103/PhysRevLett.101.191601
  [arXiv:0803.3483 [hep-th]].
  %%CITATION = doi:10.1103/PhysRevLett.101.191601;%%
  %167 citations counted in INSPIRE as of 18 May 2016

%\cite{Nakano:2008xc}
\bibitem{Nakano:2008xc} 
  E.~Nakano and W.~Y.~Wen,
  %``Critical magnetic field in a holographic superconductor,''
  Phys.\ Rev.\ D {\bf 78}, 046004 (2008)
  doi:10.1103/PhysRevD.78.046004
  [arXiv:0804.3180 [hep-th]].
  %%CITATION = doi:10.1103/PhysRevD.78.046004;%%
  %114 citations counted in INSPIRE as of 18 May 2016

%\cite{Albash:2008eh}
\bibitem{Albash:2008eh} 
  T.~Albash and C.~V.~Johnson,
  %``A Holographic Superconductor in an External Magnetic Field,''
  JHEP {\bf 0809}, 121 (2008)
  doi:10.1088/1126-6708/2008/09/121
  [arXiv:0804.3466 [hep-th]].
  %%CITATION = doi:10.1088/1126-6708/2008/09/121;%%
  %137 citations counted in INSPIRE as of 18 May 2016

%\cite{Hartnoll:2008kx}
\bibitem{Hartnoll:2008kx} 
  S.~A.~Hartnoll, C.~P.~Herzog and G.~T.~Horowitz,
  %``Holographic Superconductors,''
  JHEP {\bf 0812}, 015 (2008)
  doi:10.1088/1126-6708/2008/12/015
  [arXiv:0810.1563 [hep-th]].
  %%CITATION = doi:10.1088/1126-6708/2008/12/015;%%
  %656 citations counted in INSPIRE as of 18 May 2016

%\cite{Kruczenski:2003uq}
\bibitem{Kruczenski:2003uq} 
  M.~Kruczenski, D.~Mateos, R.~C.~Myers and D.~J.~Winters,
  %``Towards a holographic dual of large N(c) QCD,''
  JHEP {\bf 0405}, 041 (2004)
  doi:10.1088/1126-6708/2004/05/041
  [hep-th/0311270].
  %%CITATION = doi:10.1088/1126-6708/2004/05/041;%%
  %389 citations counted in INSPIRE as of 19 May 2016

%\cite{Sakai:2004cn}
\bibitem{Sakai:2004cn} 
  T.~Sakai and S.~Sugimoto,
  %``Low energy hadron physics in holographic QCD,''
  Prog.\ Theor.\ Phys.\  {\bf 113}, 843 (2005)
  doi:10.1143/PTP.113.843
  [hep-th/0412141].
  %%CITATION = doi:10.1143/PTP.113.843;%%
  %979 citations counted in INSPIRE as of 19 May 2016

%\cite{Erlich:2005qh}
\bibitem{Erlich:2005qh} 
  J.~Erlich, E.~Katz, D.~T.~Son and M.~A.~Stephanov,
  %``QCD and a holographic model of hadrons,''
  Phys.\ Rev.\ Lett.\  {\bf 95}, 261602 (2005)
  doi:10.1103/PhysRevLett.95.261602
  [hep-ph/0501128].
  %%CITATION = doi:10.1103/PhysRevLett.95.261602;%%
  %745 citations counted in INSPIRE as of 19 May 2016

%\cite{DaRold:2005mxj}
\bibitem{DaRold:2005mxj} 
  L.~Da Rold and A.~Pomarol,
  %``Chiral symmetry breaking from five dimensional spaces,''
  Nucl.\ Phys.\ B {\bf 721}, 79 (2005)
  doi:10.1016/j.nuclphysb.2005.05.009
  [hep-ph/0501218].
  %%CITATION = doi:10.1016/j.nuclphysb.2005.05.009;%%
  %554 citations counted in INSPIRE as of 19 May 2016

%\cite{Aharony:2006da}
\bibitem{Aharony:2006da} 
  O.~Aharony, J.~Sonnenschein and S.~Yankielowicz,
  %``A Holographic model of deconfinement and chiral symmetry restoration,''
  Annals Phys.\  {\bf 322}, 1420 (2007)
  doi:10.1016/j.aop.2006.11.002
  [hep-th/0604161].
  %%CITATION = doi:10.1016/j.aop.2006.11.002;%%
  %242 citations counted in INSPIRE as of 20 May 2016

%\cite{Herzog:2006ra}
\bibitem{Herzog:2006ra} 
  C.~P.~Herzog,
  %``A Holographic Prediction of the Deconfinement Temperature,''
  Phys.\ Rev.\ Lett.\  {\bf 98}, 091601 (2007)
  doi:10.1103/PhysRevLett.98.091601
  [hep-th/0608151].
  %%CITATION = doi:10.1103/PhysRevLett.98.091601;%%
  %149 citations counted in INSPIRE as of 20 May 2016

%\cite{Hawking:1982dh}
\bibitem{Hawking:1982dh} 
  S.~W.~Hawking and D.~N.~Page,
  %``Thermodynamics of Black Holes in anti-De Sitter Space,''
  Commun.\ Math.\ Phys.\  {\bf 87}, 577 (1983).
  doi:10.1007/BF01208266
  %%CITATION = doi:10.1007/BF01208266;%%
  %1129 citations counted in INSPIRE as of 21 May 2016

\bibitem{Nambu}
Y.~Nambu, 
Phys.\ Rev.\ {\bf D10}, (1974) 4262

\bibitem{tHooft}
G.~ 't Hooft, in "High Energy Physics", Proceedings of the EPS International Conference, Palermo 1975, ed. A. Zichichi, Bologna 1976.

\bibitem{Mandelstam}
 S.~ Mandelstam,
Phys.\ Rept.\ {\bf 23}, (1976) 245.

%\cite{Chen:2009pt}
\bibitem{Chen:2009pt} 
  J.~W.~Chen, Y.~J.~Kao and W.~Y.~Wen,
  %``Peak-Dip-Hump from Holographic Superconductivity,''
  Phys.\ Rev.\ D {\bf 82}, 026007 (2010)
  doi:10.1103/PhysRevD.82.026007
  [arXiv:0911.2821 [hep-th]].
  %%CITATION = doi:10.1103/PhysRevD.82.026007;%%
  %42 citations counted in INSPIRE as of 21 May 2016

%\cite{Faulkner:2009am}
\bibitem{Faulkner:2009am} 
  T.~Faulkner, G.~T.~Horowitz, J.~McGreevy, M.~M.~Roberts and D.~Vegh,
  %``Photoemission 'experiments' on holographic superconductors,''
  JHEP {\bf 1003}, 121 (2010)
  doi:10.1007/JHEP03(2010)121
  [arXiv:0911.3402 [hep-th]].
  %%CITATION = doi:10.1007/JHEP03(2010)121;%%
  %76 citations counted in INSPIRE as of 21 May 2016

%\cite{Gubser:2009dt}
\bibitem{Gubser:2009dt} 
  S.~S.~Gubser, F.~D.~Rocha and P.~Talavera,
  %``Normalizable fermion modes in a holographic superconductor,''
  JHEP {\bf 1010}, 087 (2010)
  doi:10.1007/JHEP10(2010)087
  [arXiv:0911.3632 [hep-th]].
  %%CITATION = doi:10.1007/JHEP10(2010)087;%%
  %49 citations counted in INSPIRE as of 21 May 2016

%\cite{Horowitz:2008bn}
\bibitem{Horowitz:2008bn} 
  G.~T.~Horowitz and M.~M.~Roberts,
  %``Holographic Superconductors with Various Condensates,''
  Phys.\ Rev.\ D {\bf 78}, 126008 (2008)
  doi:10.1103/PhysRevD.78.126008
  [arXiv:0810.1077 [hep-th]].
  %%CITATION = doi:10.1103/PhysRevD.78.126008;%%
  %236 citations counted in INSPIRE as of 22 May 2016

%\cite{Hofmann:2005dt}
\bibitem{Hofmann:2005dt} 
  R.~Hofmann,
  %``Nonperturbative approach to Yang-Mills thermodynamics,''
  Int.\ J.\ Mod.\ Phys.\ A {\bf 20}, 4123 (2005)
  doi:10.1142/S0217751X06035129, 10.1142/S0217751X05023931
  [hep-th/0504064].
  %%CITATION = doi:10.1142/S0217751X06035129, 10.1142/S0217751X05023931;%%
  %54 citations counted in INSPIRE as of 17 Jul 2016

%\cite{Faulkner:2009wj}
\bibitem{Faulkner:2009wj} 
  T.~Faulkner, H.~Liu, J.~McGreevy and D.~Vegh,
  %``Emergent quantum criticality, Fermi surfaces, and AdS(2),''
  Phys.\ Rev.\ D {\bf 83}, 125002 (2011)
  doi:10.1103/PhysRevD.83.125002
  [arXiv:0907.2694 [hep-th]].
  %%CITATION = doi:10.1103/PhysRevD.83.125002;%%
  %370 citations counted in INSPIRE as of 22 Jul 2016

%\cite{Kaplan:2009kr}
\bibitem{Kaplan:2009kr} 
  D.~B.~Kaplan, J.~W.~Lee, D.~T.~Son and M.~A.~Stephanov,
  %``Conformality Lost,''
  Phys.\ Rev.\ D {\bf 80}, 125005 (2009)
  doi:10.1103/PhysRevD.80.125005
  [arXiv:0905.4752 [hep-th]].
  %%CITATION = doi:10.1103/PhysRevD.80.125005;%%
  %164 citations counted in INSPIRE as of 22 Jul 2016

%\cite{Wen:2010et}
\bibitem{Wen:2010et} 
  W.~Y.~Wen,
  %``Quantum criticality in Einstein–Maxwell-dilaton gravity,''
  Phys.\ Lett.\ B {\bf 707}, 398 (2012)
  doi:10.1016/j.physletb.2011.12.053
  [arXiv:1009.3952 [hep-th]].
  %%CITATION = doi:10.1016/j.physletb.2011.12.053;%%
  %4 citations counted in INSPIRE as of 22 Jul 2016

%\cite{Unsal:2007jx}
\bibitem{Unsal:2007jx} 
  M.~Unsal,
  %``Magnetic bion condensation: A New mechanism of confinement and mass gap in four dimensions,''
  Phys.\ Rev.\ D {\bf 80}, 065001 (2009)
  doi:10.1103/PhysRevD.80.065001
  [arXiv:0709.3269 [hep-th]].
  %%CITATION = doi:10.1103/PhysRevD.80.065001;%%
  %94 citations counted in INSPIRE as of 23 Jul 2016


\end{thebibliography}
\end{document}